\documentclass[amsmath, pra, twocolumn, showpacs, floatfix]{revtex4-1}
\usepackage{graphicx}
\usepackage{dsfont}

\begin{document}   

\title{Direction-dependent coupling between a nanofiber-guided light field and a two-level atom with an electric quadrupole transition}
 
\author{Fam Le Kien,$^{1}$ S\'{i}le Nic Chormaic,$^{2}$ and Thomas Busch$^{1}$}

\affiliation{$^1$Quantum Systems Unit, Okinawa Institute of Science and Technology Graduate University, Onna, Okinawa 904-0495, Japan\\
	$^2$Light-Matter Interactions Unit, Okinawa Institute of Science and Technology Graduate University, Onna, Okinawa 904-0495, Japan
}

\date{\today}

\begin{abstract}
We study the directional dependence of the coupling between a nanofiber-guided light field and a two-level atom with an electric quadrupole transition. We examine the situation where the atom lies on the fiber transverse axis $x$, 
the quantization axis for the atomic internal states is the other orthogonal transverse axis $y$, the atomic upper and lower levels are the magnetic sublevels $M'$ and $M$ of hyperfine-structure levels of an alkali-metal atom,
and the field is in a quasilinearly polarized fundamental guided mode HE$_{11}$ with the polarization $\xi=x$ or $y$. 
We find that the absolute value of the quadrupole Rabi frequency depends on the propagation direction of the light field in the cases of ($M'-M=\pm1$, $\xi=y$) and ($M'-M=\pm2$, $\xi=x$).
We show that the directional dependence of the coupling leads to the directional dependence of spontaneous emission into  guided modes. We find that the directional dependence of the atom-field coupling in the case of quadrupole transitions is not entirely due to spin-orbit coupling of light: there are some other contributions resulting from the gradient of the spatial phase factor of the field.
\end{abstract}

\maketitle

\section{Introduction}
\label{sec:intro}

It is known that when an atom with a rotating electric dipole interacts with a light field confined in a mode of a macroscopic body, such as a nanofiber \cite{Petersen2014,Mitsch14b,Fam2014,sponhigh,Scheel2015,Jacob2016}, 
a flat surface \cite{Jacob2016,flat,Buhmann2018}, a photonic topological material \cite{Anzetta2018a,Anzetta2018b}, 
a photonic crystal waveguide \cite{leFeberScience2015}, and a nonreciprocal medium \cite{Fuchs2017}, the strength of the atom-field coupling may become asymmetric with respect to opposite propagation directions of the field. This chiral effect is due to the existence of a nonzero longitudinal field component, which oscillates in phase quadrature with respect to a nonzero transverse field component and hence creates a local transverse spin angular momentum. Due to the time reversal symmetry, a reverse of the propagation direction leads to a change in the sign of
the local transverse spin, that is, the local transverse spin is locked to the propagation direction \cite{Zeldovich,Bliokh review,Bliokh2014,Bliokh review2015,Bliokh2015,Banzer review2015,Lodahl2017}. Thus, the directional dependence of coupling between a confined light field and an atom with a rotating electric dipole is a result of spin-orbit coupling of light carrying transverse spin angular momentum  \cite{Zeldovich,Bliokh review,Bliokh2014,Bliokh review2015,Bliokh2015,Banzer review2015,Lodahl2017}.

Electric quadrupole transitions have been studied for atoms in free space \cite{Nilsen1977,Nilsen1978,Freedhoff1989,James1998,Kaler16,Afanasev2016,Peshkov2017,Afanasev2017a,Afanasev2017b,BDeb14,Ducloy16,Willitsch14,Gould09}, in evanescent fields \cite{Tojo2004,Tojo2005a,Tojo2005b}, near  dielectric microspheres \cite{Klimov1996}, near ideally conducting cylinders \cite{Klimov2000}, near plasmonic nanostructures \cite{Plasmon12,Shibata2017}, and near nanofibers \cite{quadrupole,Ray2020}. 
Recently, excitations of electric quadrupole transitions of alkali-metal atoms using nanofiber-guided light fields have been experimentally realized \cite{Ray2020}. 
Unlike electric dipole transitions, electric quadrupole transitions depend on the gradients of the field components. Furthermore, the structure of the quadrupole tensor is more complicated than that of the dipole vector. Consequently, the directional dependence of the coupling between a confined light field and an atom with a quadrupole transition is not simple and deeper insight into the involved processes is desirable. 

The aim of this paper is to study the directional dependence of the coupling between a nanofiber-guided light field and a two-level atom with an electric quadrupole transition.
We investigate the situation where the atom lies on the fiber transverse axis $x$, 
the quantization axis for the atomic internal states is the other orthogonal transverse axis $y$,
the atomic upper and lower levels are the magnetic sublevels $M'$ and $M$ of hyperfine-structure (hfs) levels of an alkali-metal atom,
and the field is in a quasilinearly polarized fundamental guided mode HE$_{11}$ with the polarization $\xi=x$ or $y$. 
We find that the absolute value of the quadrupole Rabi frequency depends on the propagation direction of the field in the cases where the atomic internal states and the polarization of the field are appropriate. We show that the  directional dependence of the atom-field coupling in the case of quadrupole transitions is partly but not entirely due to spin-orbit coupling of light. 

The paper is organized as follows. In Sec.~\ref{sec:model}, we describe the model of a two-level atom with an electric quadrupole transition driven by a guided light field of an optical nanofiber.
In Sec.~\ref{sec:Rabi}, we study the directional dependence of the coupling between the atom and the nanofiber-guided light field. In Sec.~\ref{sec:num}, we present the results of numerical calculations for the quadrupole 
Rabi frequency and the asymmetry parameter. Our conclusions are given in Sec.~\ref{sec:summary}.

\section{Model}
\label{sec:model}

We consider a two-level atom with an electric quadrupole transition interacting with a guided light field of a nearby optical nanofiber (see Fig.~\ref{fig1}). We review the descriptions of the atomic electric quadrupole and the nanofiber-guided light field below.

\subsection{Electric quadrupole transition between two magnetic levels of an alkali-metal atom}

\begin{figure}[tbh]
\begin{center}
  \includegraphics{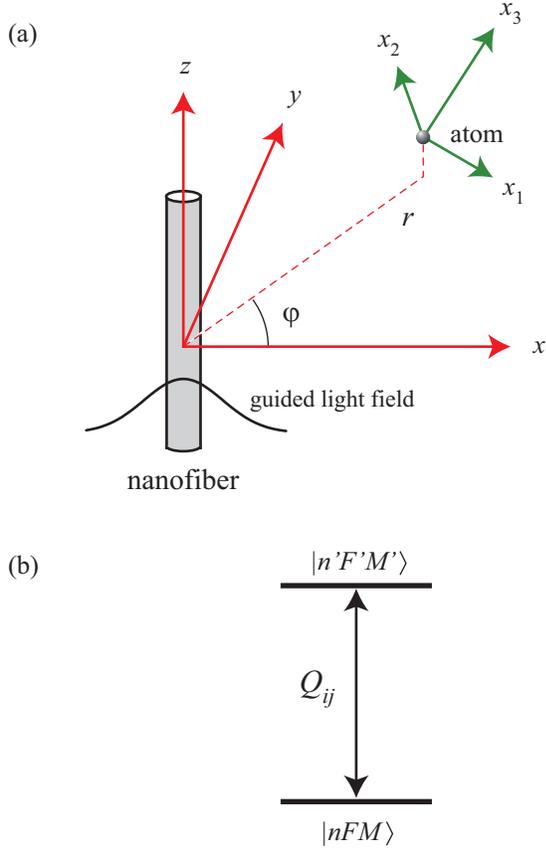}
 \end{center}
\caption{ 
(a) Atom with the local quantization coordinate system $\{x_1,x_2,x_3\}$  
in the vicinity of an optical nanofiber with
the fiber-based Cartesian coordinate system $\{x,y,z\}$ and the corresponding cylindrical coordinate system $\{r,\varphi,z\}$.
(b) Schematic of a two-level atom with an electric quadrupole transition. The upper  and lower levels of the atom are the magnetic sublevels $|n'F'M'\rangle$ and $|nFM\rangle$, respectively, of hfs levels of a realistic alkali-metal atom. The transition between the two levels is characterized by the electric quadrupole tensor $Q_{ij}$ with $i,j=1,2,3$. 
}
\label{fig1}
\end{figure}

We assume that the atom under consideration has a single valence electron. To describe the electric quadrupole and the internal states of the atom, we use the local Cartesian coordinate system $\{x_1,x_2,x_3\}$, where the origin $\mathbf{x}=0$ is located at the position of the center of mass of the atom [see Fig.~\ref{fig1}(a)]. The electric quadrupole moment tensor of the atom is given as \cite{Jackson}
\begin{equation} 
Q_{ij}=e(3x_ix_j-R^2\delta_{ij})
\end{equation} 
for $i,j=1,2,3$, where $x_i$ and $x_j$ are the $i$th and $j$th coordinates of the valence electron and $R=\sqrt{x_1^2+x_2^2+x_3^2}$ is the distance from the electron to the center of mass of the atom. 

We assume that the optical driving field is near to resonance with a quadrupole transition between two atomic internal states, namely the upper state $|e\rangle$ with the  energy $\hbar\omega_e$ and the lower state $|g\rangle$ with the energy $\hbar\omega_g$. 
We present the electric component $\mathbf{E}$ of the optical field in the form $\mathbf{E}=(\boldsymbol{\mathcal{E}}e^{-i\omega t}+\boldsymbol{\mathcal{E}}^\ast e^{i\omega t})/2$,
where $\boldsymbol{\mathcal{E}}$ is the field amplitude and $\omega$ is the field frequency. 
The interaction Hamiltonian of the system in the interaction picture and the rotating-wave approximation reads
\begin{equation}\label{q4}
	H_I=-\frac{\hbar}{2}\Omega e^{-i(\omega-\omega_0) t}\sigma_{eg}+\mathrm{H.c.},
\end{equation}
where $\omega_0=\omega_e-\omega_g$ is the atomic transition frequency,  $\sigma_{eg}=|e\rangle\langle g|$ is the atomic transition operator, and
\begin{equation}\label{q5}
	\Omega=\frac{1}{6\hbar}\sum_{ij}\langle e|Q_{ij}|g\rangle\frac{\partial \mathcal{E}_j}{\partial x_i}(0)
\end{equation}
is the Rabi frequency for the quadrupole transition between the levels $|e\rangle$ and $|g\rangle$  \cite{James1998}. 
In Eq.~(\ref{q5}), the spatial derivatives of the field components  $\mathcal{E}_j$ with respect to the coordinates $x_i$ are calculated at the position $\mathbf{x}=0$ of the atom. 

To be concrete, we consider the quadrupole transition between the magnetic sublevels $|e\rangle=|n'F'M'\rangle$ and $|g\rangle=|nFM\rangle$  of an alkali-metal atom [see Fig.~\ref{fig1}(b)]. 
Here, $n'$ and $n$ denote the principal quantum numbers and also all additional quantum numbers not shown explicitly, 
$F'$ and $F$ are the quantum numbers for the total angular momenta of the atomic internal states, and $M'$ and $M$ are the magnetic quantum numbers. 
The matrix elements $\langle n'F'M'|Q_{ij}|nFM\rangle$ of the quadrupole operators $Q_{ij}$ are given as \cite{James1998,quadrupole}
\begin{eqnarray}\label{q2}
\lefteqn{\langle n'F'M'|Q_{ij}|nFM\rangle=3e u_{ij}^{(M'-M)}(-1)^{F'-M'}}\nonumber\\
&&\mbox{}\times
\begin{pmatrix}F' &2 &F \\-M' & M'-M& M\end{pmatrix}
\langle n'F'\|T^{(2)}\|nF\rangle,\qquad
\end{eqnarray}
where the matrices $u_{ij}^{(q)}$ with $i,j=1,2,3$ and $q=M'-M=-2,-1,0,1,2$ characterize the structures of the spherical components of the quadrupole tensor $Q_{ij}$ and are given as
\begin{eqnarray}\label{q3}
u_{ij}^{(0)}&=&\frac{1}{\sqrt6}\begin{pmatrix} -1&0&0\\0&-1&0\\0&0&2\end{pmatrix},\nonumber\\
u_{ij}^{(\pm1)}&=&\frac{1}{2}\begin{pmatrix} 0&0&\mp1\\0&0&i\\\mp1&i&0\end{pmatrix},\nonumber\\
u_{ij}^{(\pm2)}&=&\frac{1}{2}\begin{pmatrix} 1&\mp i&0\\\mp i&-1&0\\0&0&0\end{pmatrix}.
\end{eqnarray}
In Eq.~(\ref{q2}), the array in the parentheses is a 3$j$ symbol and the invariant factor $\langle n'F' \| T^{(2)}\|nF \rangle$ is the reduced matrix element of the tensor operators $T_{q}^{(2)}=2(2\pi/15)^{1/2} R^2Y_{2q}(\vartheta,\phi)$. Here, $Y_{lq}$ is a spherical harmonic function of degree $l$ and order $q$,
and $\vartheta$ and $\phi$ are spherical angles in the spherical coordinates $\{R,\vartheta,\phi\}$ associated with the local Cartesian coordinates $\{x_1,x_2,x_3\}$.

When we insert Eq.~\eqref{q2} into Eq.~\eqref{q5}, we obtain \cite{James1998}
\begin{equation}\label{q6}
\Omega=C_{F'M'FM}S_{M'-M},
\end{equation}
where 
\begin{eqnarray}\label{q6a}
C_{F'M'FM}&=&\frac{e}{2\hbar}(-1)^{F'-M'}
\begin{pmatrix}F' &2 &F \\-M' & M'-M& M\end{pmatrix} 
\nonumber\\&&
\times\langle n'F'\|T^{(2)}\|nF\rangle
\end{eqnarray}
is a proportionality coefficient and
\begin{equation}\label{q6b}
S_{M'-M}=\sum_{ij}u_{ij}^{(M'-M)}\frac{\partial \mathcal{E}_j}{\partial x_i}(0)
\end{equation}
is a reduced coupling factor. Note that $C_{F'M'FM}$ depends on the atom but not on the field and $S_{M'-M}$
depends on the difference $M'-M$ but not on $M'$ and $M$ separately.

The electric quadrupole transition selection rules for $F$ and $F'$ and for $M$ and $M'$ are $|F'-F|\le 2\le F'+F$ and $|M'-M|\le 2$. For the quantum numbers $J$ and $J'$ of the total electronic angular momenta, the selection rules are $|J'-J|\le 2\le J'+J$. For the quantum numbers $L$ and $L'$ of the orbital electronic angular momenta, the selection rules read $|L'-L|=0,2$ and $L'+L\ge 2$. 

For a plane-wave light field $\boldsymbol{\mathcal{E}}=\mathcal{E}_0\boldsymbol{\epsilon}e^{ i\mathbf{k}\cdot\mathbf{x}}$ with the  amplitude $\mathcal{E}_0$, the polarization vector $\boldsymbol{\epsilon}$, and the wave vector $\mathbf{k}$ in free space, the reduced coupling factor is found to be $S_q= i\mathcal{E}_0(\mathbf{k}\cdot\mathbf{u}^{(q)}\cdot\boldsymbol{\epsilon})$. For such a field, $|S_q|$ and hence $|\Omega|$ do not change when the direction of the wave vector $\mathbf{k}$ is reversed. Thus, the strength of the coupling between a plane-wave light field and an atom with a quadrupole transition is symmetric with respect to the opposite propagation directions. We will show in the next section that the directional symmetry of coupling is not valid in the case of quasilinearly polarized nanofiber-guided light fields. 

\subsection{Quasilinearly polarized nanofiber-guided field}

We assume that the external field interacting with the atom is the guided light field of a nearby optical nanofiber [see Fig.~\ref{fig1}(a)] \cite{TongNat03,review2016,review2017,review2018}. The fiber is a dielectric cylinder of radius $a$ and refractive index $n_1$ and is surrounded by an infinite background medium of refractive index $n_2$, where $n_2<n_1$. To describe the guided field, we use Cartesian coordinates $\{x,y,z\}$, where $z$ lies along the fiber axis, and also cylindrical coordinates $\{r,\varphi,z\}$, where $r$ and $\varphi$ are the polar coordinates in the cross-sectional plane $xy$. 

We study the case of a single-mode vacuum-clad nanofiber where $n_2=1$ and the fiber radius is small enough so that it can support only the fundamental guided mode  HE$_{11}$  
in a finite bandwidth around the central frequency $\omega_0=\omega_e-\omega_g$ of the atom \cite{TongNat03,review2016,review2017,review2018}. The single-mode condition for the fiber reads $ka\sqrt{n_1^2-n_2^2}<2.405$,
where $k=\omega/c$ is the wave number of the light field in free space \cite{fiber books}.
The theory of guided modes of cylindrical fibers is described in Ref.~\cite{fiber books} and is summarized and analyzed in detail for nanofibers in Ref.~\cite{highorder}. 

We assume that the field is prepared in a quasilinearly polarized fundamental guided mode $\mathrm{HE}_{11}$ \cite{fiber books,highorder}.
The amplitude of the electric part of the field in the mode is \cite{fiber books,highorder}
\begin{eqnarray}\label{q7}
	\boldsymbol{\mathcal{E}}&=&\mathcal{A}[\hat{\mathbf{r}}e_r\cos (\varphi-\varphi_0)
	+i\hat{\boldsymbol{\varphi}}e_\varphi\sin (\varphi-\varphi_0)
	\nonumber\\&&\mbox{} 	
	+f\hat{\mathbf{z}}e_z\cos (\varphi-\varphi_0)]e^{if\beta z}.
\end{eqnarray}
Here, $\beta>0$ is the longitudinal propagation constant, $f=+1$ or $-1$ denotes the forward or backward propagation direction along the fiber axis $z$, and $\varphi_0$ is the azimuthal orientation angle for the principal polarization axis in the fiber transverse plane $xy$.  
The profile functions $e_r=e_r(r)$, $e_\varphi=e_\varphi(r)$, and $e_z=e_z(r)$ are the cylindrical components of 
the fundamental guided mode with the forward propagation direction and the counterclockwise quasicircular polarization,
depend on $r$ but not on $\varphi$ and $z$, and are given in Ref.~\cite{fiber books,highorder}. 
The constant $\mathcal{A}$ depends on the mode power. 
The  relative phases between $e_r$ and $e_\varphi$ and between $e_r$ and $e_z$ are $\pm\pi/2$ \cite{fiber books,highorder}.
For an appropriate choice of a common phase factor for the mode profile functions, we have \cite{fiber books,highorder}
\begin{equation}\label{7a}
	e^*_r=-e_r,\quad e^*_{\varphi}=e_{\varphi},\quad e^*_z=e_z,
\end{equation}
that is, $e_r$ is purely imaginary and $e_\varphi$ and $e_z$ are purely real. 
It follows from Eqs.~(\ref{q7}) and (\ref{7a}) that $\boldsymbol{\mathcal{E}}|_{f=+1}\propto \boldsymbol{\mathcal{E}}^*|_{f=-1}$. This relation indicates that the quasilinearly polarized modes having the opposite propagation directions $f=\pm1$ and the same polarization orientation angle $\varphi_0$ are the time reversal of each other.
For $\varphi_0=0$ or $\pi/2$, the mode is quasilinearly polarized along the $x$ or $y$ direction, respectively.

In Eq.~(\ref{q7}), the mode profile function $e_z$ for the longitudinal component of the field is accompanied by the factor $f=+1$ or $-1$, which corresponds to the forward or backward propagation direction of light, respectively. This directional dependence is a consequence of the time reversal symmetry and leads to the spin-orbit coupling of light carrying transverse spin angular momentum
\cite{Zeldovich,Bliokh review,Bliokh review2015,Bliokh2014,Bliokh2015}.
Indeed, the spin angular momentum density of light in the Abraham formulation is given by $\mathbf{j}^{\mathrm{(sp)}}=(\epsilon_0/2\omega)\mathrm{Im}(\boldsymbol{\mathcal{E}}^*\times \boldsymbol{\mathcal{E}})$.
From Eqs.~(\ref{q7}) and (\ref{7a}), we find $\mathbf{j}^{\mathrm{(sp)}}=j_r^{\mathrm{(sp)}}\hat{\mathbf{r}}+j_\varphi^{\mathrm{(sp)}}\hat{\boldsymbol{\varphi}}$, where
$j_r^{\mathrm{(sp)}}=-(\epsilon_0|\mathcal{A}|^2/\omega)f\sin(\varphi-\varphi_0)\cos(\varphi-\varphi_0) \mathrm{Re}(e_\varphi e_z^*)$
and $j_\varphi^{\mathrm{(sp)}}=(\epsilon_0|\mathcal{A}|^2/\omega)f\cos^2(\varphi-\varphi_0) \mathrm{Im}(e_re_z^*)$.
It is clear that the local spin vector $\mathbf{j}^{\mathrm{(sp)}}$ lies in the transverse plane $xy$ and flips with the reversion of the field propagation direction $f$.

Note that the basis unit vectors 
$\hat{\mathbf{r}}=\cos\varphi\,\hat{\mathbf{x}}+\sin\varphi\,\hat{\mathbf{y}}$ and $\hat{\boldsymbol{\varphi}}=-\sin\varphi\,\hat{\mathbf{x}}+\cos\varphi\,\hat{\mathbf{y}}$ depend on the azimuthal angle $\varphi$. In the cylindrical coordinates $\{r,\varphi,z\}$, the Cartesian components $\mathcal{E}_x$, $\mathcal{E}_y$, and $\mathcal{E}_z$ of the field in a quasilinearly polarized fundamental guided mode are found from Eq.~(\ref{q7}) to be
\begin{eqnarray}\label{q8}
	\mathcal{E}_x&=&\mathcal{A}[e_r\cos (\varphi-\varphi_0)\cos\varphi
	-ie_\varphi\sin (\varphi-\varphi_0)\sin\varphi]
	\nonumber\\&&\mbox{}
	\times e^{if\beta z},
	\nonumber\\
	\mathcal{E}_y&=&\mathcal{A}[e_r\cos (\varphi-\varphi_0)\sin\varphi
	+ie_\varphi\sin (\varphi-\varphi_0)\cos\varphi]
	\nonumber\\&&\mbox{}
	\times e^{if\beta z},
	\nonumber\\
	\mathcal{E}_z&=&\mathcal{A} fe_z\cos (\varphi-\varphi_0)e^{if\beta z}.
\end{eqnarray}
When we set $\varphi_0=0$, we find from Eqs.~(\ref{q8}) the following expressions for 
the Cartesian components of the $x$-polarized guided field $\boldsymbol{\mathcal{E}}^{(fx)}$:
\begin{eqnarray}\label{q8a}
	\mathcal{E}_x^{(fx)}&=&\mathcal{A}(e_r\cos^2\varphi
	-ie_\varphi\sin^2\varphi) e^{if\beta z},
	\nonumber\\
	\mathcal{E}_y^{(fx)}&=&\mathcal{A}(e_r+ie_\varphi)\sin\varphi\cos\varphi\; e^{if\beta z},
	\nonumber\\
	\mathcal{E}_z^{(fx)}&=&\mathcal{A} fe_z\cos\varphi\; e^{if\beta z}.
\end{eqnarray}
For $\varphi_0=\pi/2$, Eqs.~(\ref{q8}) yield the following expressions for
the Cartesian components of the $y$-polarized guided field $\boldsymbol{\mathcal{E}}^{(fy)}$:  
\begin{eqnarray}\label{q8b}
	\mathcal{E}_x^{(fy)}&=&\mathcal{A}(e_r+ie_\varphi)\sin\varphi\cos\varphi\; e^{if\beta z},
	\nonumber\\
	\mathcal{E}_y^{(fy)}&=&\mathcal{A}(e_r\sin^2\varphi
	-ie_\varphi\cos^2\varphi)e^{if\beta z},
	\nonumber\\
	\mathcal{E}_z^{(fy)}&=&\mathcal{A} fe_z\sin\varphi\; e^{if\beta z}.
\end{eqnarray}

\section{Directional dependence of the atom-field coupling}
\label{sec:Rabi}

We show in this section that the interaction between a quasilinearly polarized nanofiber-guided light field and an atom with a quadrupole transition may depend on the propagation direction $f$. We demonstrate analytically that the directional dependence of coupling occurs when the polarization of the field and 
the internal states of the atom, characterized by the orientation of the quantization axis and the magnetic quantum numbers,  are appropriate. 

First, we show that
in the case where the quantization axis is the fiber axis $z$, that is, $x_3\parallel z$, the absolute value $|\Omega|$ of the Rabi frequency of the quadrupole transition does not depend on the propagation direction $f$. 
Indeed, for  $x_1\parallel x$, $x_2\parallel y$, and $x_3\parallel z$, Eq.~(\ref{q6b}) for the reduced coupling factors $S_q$ with $q=M'-M=0,\pm1,\pm2$ yields the following expressions:
\begin{eqnarray}\label{q9a}
	S_0&=&
	\frac{1}{\sqrt6}\bigg(-\frac{\partial \mathcal{E}_x}{\partial x}
	-\frac{\partial \mathcal{E}_y}{\partial y}
	+2\frac{\partial \mathcal{E}_z}{\partial z}\bigg),
	\nonumber\\
	S_{\pm1}&=&
	\frac{1}{2}\bigg(\mp\frac{\partial \mathcal{E}_z}{\partial x}
	\mp\frac{\partial \mathcal{E}_x}{\partial z}
	+i\frac{\partial \mathcal{E}_z}{\partial y}
	+i\frac{\partial \mathcal{E}_y}{\partial z}\bigg),
	\nonumber\\
	S_{\pm2}&=&
	\frac{1}{2}\bigg(\frac{\partial \mathcal{E}_x}{\partial x}
	-\frac{\partial \mathcal{E}_y}{\partial y}
	\mp i\frac{\partial \mathcal{E}_y}{\partial x}
	\mp i\frac{\partial \mathcal{E}_x}{\partial y}\bigg).
\end{eqnarray}
According to Eqs.~(\ref{q8}), $\mathcal{E}_x$  and $\mathcal{E}_y$ depend on $f$ through the spatial phase factor $e^{if\beta z}$ and $\mathcal{E}_z$ depends on $f$ through the combined factor $fe^{if\beta z}$, which is the product of $f$ and $e^{if\beta z}$. Then, it follows from Eqs.~(\ref{q9a}) that $S_0$ and $S_{\pm2}$ depend on $f$ through the phase factor $e^{if\beta z}$ and $S_{\pm1}$ depends on $f$ through the factor $fe^{if\beta z}$. Hence, it follows from Eq.~(\ref{q6}) that the absolute value $|\Omega|$ of the Rabi frequency for the quadrupole transition between the magnetic sublevels does not depend on $f$ in the case where the quantization axis is the fiber axis $z$. 

Next, we show that $|\Omega|$ depends on $f$ 
in the general case of interaction between a quasilinearly polarized nanofiber-guided light field and an atom with a quadrupole transition. We demonstrate the $f$ dependence of $|\Omega|$ in a particular case where the quantization axis is a fiber transverse axis. We choose such a quantization axis for consideration because it has led to direction-dependent spontaneous emission from an atom with a dipole transition  \cite{Mitsch14b,Fam2014}. To be concrete, we use the fiber transverse axis $y$ as the quantization axis, that is, we take $x_3\parallel y$. In addition, we take $x_1\parallel z$ and $x_2\parallel x$. Then, Eq.~(\ref{q6b}) for the reduced coupling factors $S_q$ yields the following expressions:
\begin{eqnarray}\label{q9}
	S_0&=&
	\frac{1}{\sqrt6}\bigg(-\frac{\partial \mathcal{E}_z}{\partial z}
   -\frac{\partial \mathcal{E}_x}{\partial x}
   +2\frac{\partial \mathcal{E}_y}{\partial y}\bigg),
\nonumber\\
	S_{\pm1}&=&
	\frac{1}{2}\bigg(\mp\frac{\partial \mathcal{E}_y}{\partial z}
\mp\frac{\partial \mathcal{E}_z}{\partial y}
+i\frac{\partial \mathcal{E}_y}{\partial x}
+i\frac{\partial \mathcal{E}_x}{\partial y}\bigg),
\nonumber\\
	S_{\pm2}&=&
\frac{1}{2}\bigg(\frac{\partial \mathcal{E}_z}{\partial z}
-\frac{\partial \mathcal{E}_x}{\partial x}
\mp i\frac{\partial \mathcal{E}_x}{\partial z}
\mp i\frac{\partial \mathcal{E}_z}{\partial x}\bigg).
\end{eqnarray}

We assume that the atom is located outside the fiber and on the positive side of the axis $x$, which corresponds to the azimuthal angle $\varphi=0$ and is perpendicular to the quantization axis $y$.
In this case, we find the explicit expressions (\ref{q14}) for the coupling factors $S_{q}$.
In these expressions, the angle $\varphi_0$, which specified the orientation of the principal polarization axis of the field, is arbitrary.
When the guided field is polarized along the $x$ or $y$ direction, we have $\varphi_0=0$ or $\pi/2$, respectively. 
For convenience and clarity, we use the notation $S_{q}^{(f\xi)}=S_{q}$ to indicate explicitly that this coupling factor corresponds to the field with the propagation direction $f$ and the polarization $\xi=x,y$. 
Similarly, we use the notation $\Omega^{(f\xi)}_{q}=\Omega$ for the Rabi frequency to indicate that it corresponds to the case where the field propagation direction is $f$, the field polarization is $\xi=x,y$, and the difference between the magnetic quantum numbers of the upper and lower states is $M'-M=q$.
From Eqs.~(\ref{q14}), we get the expressions
\begin{eqnarray}\label{q18a}
S_0^{(fx)}&=&
-\frac{\mathcal{A}}{\sqrt6}\bigg[i\beta e_z+e'_r
-\frac{2}{r}(e_r+ie_\varphi)\bigg]e^{if\beta z},
\nonumber\\
S_0^{(fy)}&=&0,
\end{eqnarray}
\begin{eqnarray}\label{q19}
S_{\pm1}^{(fx)}&=&0,
\nonumber\\
S_{\pm1}^{(fy)}&=&
\frac{\mathcal{A}}{2}\bigg[\mp f\bigg(\beta e_\varphi  
+\frac{1}{r}e_z\bigg)+e'_\varphi
+\frac{i}{r}(e_r+ie_\varphi)\bigg]e^{if\beta z},
\nonumber\\
\end{eqnarray}
and
\begin{eqnarray}\label{q20}
S_{\pm2}^{(fx)}&=&
\frac{\mathcal{A}}{2}[i\beta e_z-e'_r
\mp f(ie'_z-\beta e_r)]e^{if\beta z},
\nonumber\\
S_{\pm2}^{(fy)}&=&0.
\end{eqnarray}
Note that $S_q^{(f\xi)}=S_{-q}^{(-f,\xi)}$.

It is clear from Eqs.~(\ref{q19}) and (\ref{q20}) and the relations (\ref{7a}) that the absolute values of the coupling factors $S_{\pm1}^{(fy)}$ and $S_{\pm2}^{(fx)}$ have different values for the different propagation directions $f=+1,-1$.  
Due to the directional dependencies of $|S_{\pm1}^{(fy)}|$ and $|S_{\pm2}^{(fx)}|$, the absolute values $|\Omega_{\pm1}^{(fy)}|$ and $|\Omega_{\pm2}^{(fx)}|$ of the corresponding Rabi frequencies are asymmetric with respect to $f$, that is, the atom-field coupling is chiral. Meanwhile, $|S_{0}^{(fx)}|$ and hence $|\Omega_{0}^{(fx)}|$ do not depend on $f$. Furthermore, $S_{0}^{(fy)}$, $S_{\pm1}^{(fx)}$, and $S_{\pm2}^{(fy)}$ are vanishing, and so are $\Omega_{0}^{(fy)}$, $\Omega_{\pm1}^{(fx)}$, and $\Omega_{\pm2}^{(fy)}$.

The asymmetry of the absolute values of the Rabi frequencies for the opposite propagation directions of the field is characterized by the parameter
\begin{equation}\label{q22}
\eta_q^{(\xi)}\equiv\frac{|\Omega_{q}^{(+,\xi)}|^2-|\Omega_{q}^{(-,\xi)}|^2}{|\Omega_{q}^{(+,\xi)}|^2+|\Omega_{q}^{(-,\xi)}|^2}
=\frac{|S_{q}^{(+,\xi)}|^2-|S_{q}^{(-,\xi)}|^2}{|S_{q}^{(+,\xi)}|^2+|S_{q}^{(-,\xi)}|^2}.
\end{equation}
Note that $|\eta_q^{(\xi)}|\leq1$.
We have $\eta_q^{(\xi)}=0$ for symmetric coupling, 
$\eta_q^{(\xi)}\not=0$ for asymmetric coupling, and $\eta_q^{(\xi)}=\pm1$ for unidirectional coupling. 
Like the coupling factor $S_q^{(f\xi)}$, the asymmetry parameter $\eta_q^{(\xi)}$ depends on $q=M'-M$ but not on $M'$ and $M$ separately.

In the cases of ($q=0$, $\xi=y$), ($q=\pm1$, $\xi=x$), or ($q=\pm2$, $\xi=y$), there is no coupling between the atom and the guided light field and, therefore, the asymmetry parameter $\eta_q^{(\xi)}$ is undefined. Meanwhile, for the cases of ($q=0$, $\xi=x$), ($q=\pm1$, $\xi=y$), or ($q=\pm2$, $\xi=x$), we find
\begin{eqnarray}\label{q18b}
	\eta_0^{(x)}&=&0,
	\nonumber\\
	\eta_{\pm1}^{(y)}&=&
	\mp 2\,\mathrm{Re}\left\{\frac{\left(\beta e_\varphi+\frac{1}{r}e_z\right)
		\left[e'_\varphi+\frac{i}{r}(e_r+ie_\varphi)\right]^*}
	{\left|\beta e_\varphi+\frac{1}{r}e_z\right|^2
		+\left|e'_\varphi+\frac{i}{r}(e_r+ie_\varphi)\right|^2}\right\},
	\nonumber\\
	\eta_{\pm2}^{(x)}&=&
	\mp 2\,\mathrm{Re}\left\{\frac{(e'_z+i\beta e_r)(\beta e_z+ie'_r)^*}
	{|e'_z+i\beta e_r|^2+|\beta e_z+ie'_r|^2}\right\}.
\end{eqnarray}
Note that $\eta_q^{(\xi)}=-\eta_{-q}^{(\xi)}$.
Equations (\ref{q18b}) show that the asymmetry parameter $\eta_q^{(\xi)}$ has not only contributions  from  the longitudinal component $e_z$ but also contributions from the transverse components $e_r$ and $e_\varphi$ of the mode profile function.

Like the case of dipole transitions \cite{Petersen2014,Mitsch14b,Fam2014},
the atom-field interaction via quadrupole transitions can be asymmetric with respect to the opposite propagation directions due to the presence of the longitudinal field component $\mathcal{E}_z\propto fe_z\cos(\varphi-\varphi_0)e^{if\beta z}$. The chiral effect caused by this field component is a signature of the
spin-orbit coupling of light carrying transverse spin angular momentum \cite{Zeldovich,Bliokh review,Bliokh review2015,Bliokh2014,Bliokh2015}. In the case of quadrupole transitions, the effects of $\mathcal{E}_z$ on the directional dependence of the atom-field coupling appear through the transverse (radial and azimuthal) gradients of this field component.

It is interesting to note that the asymmetry of the coupling with respect to the opposite propagation directions may appear for quadrupole transitions even when the function $e_z$ and hence the longitudinal field component $\mathcal{E}_z$ are vanishing. This feature is absent in the case of atoms with  dipole transitions. It appears in atoms with quadrupole transitions because the corresponding interaction between the field and the atom is proportional to a superposition of the terms associated with the gradients of the field components. Among them is the contribution of the longitudinal (axial) gradient of the spatial phase factor $e^{if\beta z}$ of the field. This phase gradient leads to the terms accompanied by the direction-dependent coefficient $f\beta$ in front of the phase factor $e^{if\beta z}$. The interference between these terms and the terms with direction-independent coefficients contribute to the directional asymmetry of the absolute value of the  Rabi frequency. It is clear that the directional dependence of the coefficient $f\beta$ in the phase gradient $if\beta e^{if\beta z}$ is due to the directional dependence of the wave vector and is related to that of the linear momentum of light. The physics of this directional dependence is not new but different from that of spin-orbit coupling of light.

It follows from the last two expressions in Eqs.~(\ref{q18b}) that, in the limit of large radial distances $r$,  we have
\begin{eqnarray}\label{q18c}
\eta_{\pm1}^{(y)}(\infty)\equiv\lim_{r\to\infty}\eta_{\pm1}^{(y)}&=&
	\pm \frac{2\beta \kappa}{\beta^2+\kappa^2},
	\nonumber\\
\eta_{\pm2}^{(x)}(\infty)\equiv\lim_{r\to\infty}\eta_{\pm2}^{(x)}&=&
	\pm \frac{4\beta \kappa(\beta^2+\kappa^2)}
	{4\beta^2 \kappa^2+(\beta^2+\kappa^2)^2}.
\end{eqnarray}
Here we have introduced the parameter $\kappa=\sqrt{\beta^2-n_2^2k^2}$. In deriving Eqs.~(\ref{q18c}) we have used the explicit expressions for the components $e_r$, $e_\varphi$, and $e_z$ of the mode profile function given in Refs.~\cite{fiber books,highorder}. It is clear
that the limiting values $\eta_{\pm1}^{(y)}(\infty)$ and $\eta_{\pm2}^{(x)}(\infty)$ of the asymmetry factors $\eta_{\pm1}^{(y)}$ and $\eta_{\pm2}^{(x)}$, respectively, are not zero although the Rabi frequency 
$\Omega$ reduces to zero with increasing $r$.

Note that in the limit of large fiber radii, we have $\beta\to kn_1$ and $\kappa\to k\sqrt{n_1^2-n_2^2}$. In this limit, the asymmetry parameters $\eta_{\pm1}^{(y)}(\infty)$ and $\eta_{\pm2}^{(x)}(\infty)$ tend to the values
\begin{eqnarray} \label{q18d}
\lim_{a\to\infty}\eta_{\pm1}^{(y)}(\infty)&=&\pm \frac{2 n_1\sqrt{n_1^2-n_2^2}}{2n_1^2-n_2^2},\nonumber\\ 
\lim_{a\to\infty}\eta_{\pm2}^{(x)}(\infty)&=&\pm\frac{ 4 n_1\sqrt{n_1^2-n_2^2}\;(2n_1^2-n_2^2)}{4 n_1^2(n_1^2-n_2^2)+(2n_1^2-n_2^2)^2}.
\end{eqnarray}
The above limiting values are also nonzero.

We now show that the directional dependence of the absolute value $|\Omega|$ of the Rabi frequency leads to the directional dependence of spontaneous emission into guided modes.
Let $\gamma_{\mathrm{g}}^{(f\xi)}$ be the rate of quadrupole spontaneous emission from the atom into the guided modes with the propagation direction $f$ and the polarization $\xi$. We show in Appendix \ref{sec:spont} that the rate $\gamma_{\mathrm{g}}^{(f\xi)}$ is proportional to  $|S_{q}^{(\mu_0)}|^2$, where $S_{q}^{(\mu)}=S_{q}|_{\boldsymbol{\mathcal{E}}=\mathbf{e}^{(\mu)}}$
is the reduced coupling factor for the normalized field in the guided mode $\mu=(\omega f\xi)$ and $\mu_0=(\omega_0f\xi)$ is the label for the guided mode at the resonant frequency. Here, the notation $\mathbf{e}^{(\mu)}$ stands for the normalized mode profile function, which is given by Eqs.~(\ref{4}) under the normalization condition (\ref{5}). 

We again assume that the quantization axis is the fiber transverse axis $y$ and the atom lies on the fiber transverse axis $x$. In this case, Eqs.~(\ref{q18a})--(\ref{q20}) are valid. It follows from these equations that the absolute value $|S_{q}^{(\mu_0)}|$ of the coupling factor for the normalized field in the mode $\mu_0=(\omega_0f\xi)$ may depend on $f$.
It is obvious that the $f$ dependence of $|S_{q}^{(\mu_0)}|$ leads to the $f$ dependence of the rate $\gamma_{\mathrm{g}}^{(f\xi)}$.

The rate of spontaneous emission into the guided modes propagating in the $f$ direction regardless of polarization is given by
\begin{equation}\label{v12}
	\gamma_{\mathrm{g}}^{(f)}=\gamma_{\mathrm{g}}^{(fx)}+\gamma_{\mathrm{g}}^{(fy)}.
\end{equation}
The asymmetry parameter for the directional dependence of the spontaneous emission rate into the nanofiber is defined as
\begin{equation}\label{v26}
	\eta_{\mathrm{g}}\equiv\frac{\gamma_{\mathrm{g}}^{(+)}-\gamma_{\mathrm{g}}^{(-)}}{\gamma_{\mathrm{g}}^{(+)}+\gamma_{\mathrm{g}}^{(-)}}.
\end{equation}
According to Eqs.~(\ref{q18a})--(\ref{q20}), the spontaneous emission rate $\gamma_{\mathrm{g}}^{(f\xi)}$ is vanishing for 
the $y$-polarized mode in the cases of $q=0$ or $\pm2$ and for the $x$-polarized mode in the cases of $q=\pm1$. Due to this fact, we have $\gamma_{\mathrm{g}}^{(f)}=\gamma_{\mathrm{g}}^{(fx)}$ for $q=0$ or $\pm2$ and $\gamma_{\mathrm{g}}^{(f)}=\gamma_{\mathrm{g}}^{(fy)}$ for $q=\pm1$. Hence, the asymmetry parameter $\eta_q$ for the directional dependence of the spontaneous emission rate into the nanofiber is found to be
\begin{equation}\label{v27}
	\eta_{\mathrm{g}}\big|_{q=0}=0,\qquad 
	\eta_{\mathrm{g}}\big|_{q=\pm1}=\eta_{\pm1}^{(y)},\qquad 
	\eta_{\mathrm{g}}\big|_{q=\pm2}=\eta_{\pm2}^{(x)},
\end{equation}
where $\eta_{\pm1}^{(y)}$ and $\eta_{\pm2}^{(x)}$ are given by Eqs.~(\ref{q18b}) with the mode profile functions $e_r$, $e_{\varphi}$, and $e_z$ being evaluated at $\omega=\omega_0$. We emphasize that the above result for the quadrupole spontaneous emission is valid only for the channel of emission into nanofiber-guided modes and only in the framework of the  model of a two-level atom. A full treatment must include the channel of emission into radiation modes and the multilevel structure of the atom.

We note that in the case where the atom lies on the quantization axis $y$, the absolute values of the coupling factors $S_q^{(f\xi)}$  and, hence,
the absolute values of the Rabi frequencies $\Omega_q^{(f\xi)}$ for the $x$- and $y$-polarized guided light fields do not depend on the field propagation direction $f$ (see Appendix \ref{sec:coupling}).

\section{Numerical results}
\label{sec:num} 

In this section, we present the results of numerical calculations for the direction-dependent coupling between a nanofiber-guided light field and a two-level atom with an electric quadrupole transition. 
As an example, we study the electric quadrupole transition between the ground state $5S_{1/2}$ 
and the excited state $4D_{5/2}$ of a $^{87}$Rb atom.
For this transition, we have $L=0$, $J=1/2$, $L'=2$, $J'=5/2$, and $I=3/2$.
The wavelength of the transition is $\lambda_0=516.5$ nm \cite{NIST}. The reduced quadrupole matrix element $\langle n'J'\|T^{(2)}\|nJ\rangle$ is deducted from the experimentally measured oscillator strength $f_{JJ'}^{(0)}=8.06\times 10^{-7}$ in free space \cite{Nilsen1978,James1998,Tojo2005b}.  
In our numerical calculations, we assume that the driving field is at exact resonance with the atom ($\omega=\omega_0$).
For most of our numerical calculations (except for Figs.~\ref{fig6} and \ref{fig7}),
we take the fiber radius $a=180$ nm, which is small enough that only the fundamental guided mode is supported.
We assume that the atom is located on the positive side of the axis $x$ and outside the fiber.

\begin{figure}[tbh]
	\begin{center}
		\includegraphics{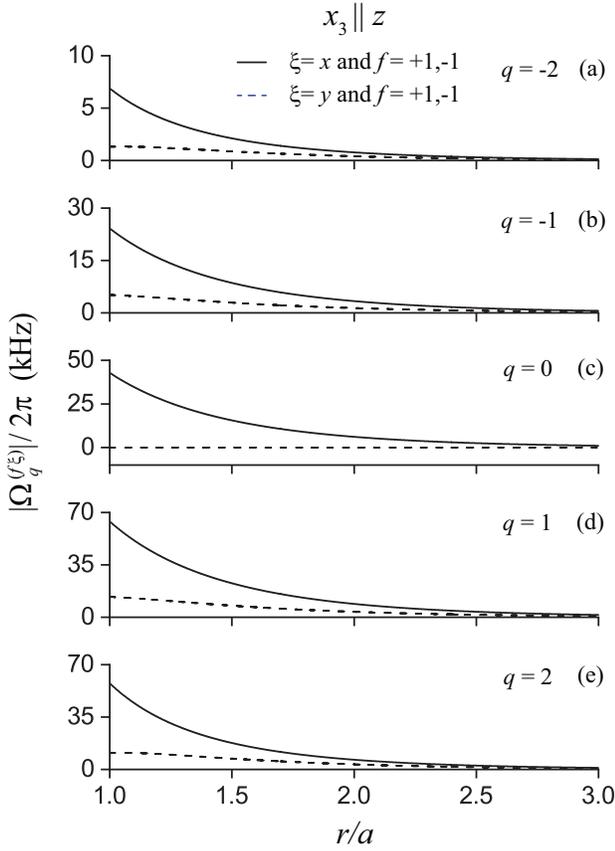}
	\end{center}
	\caption{Radial-distance dependence of the absolute value  $|\Omega_q^{(f\xi)}|$ of the Rabi frequency for the quadrupole transition between the sublevel $M=2$ of the hfs level $5S_{1/2}F=2$ and the sublevel $M'=M+q$ of the hfs level $4D_{5/2}F'=4$ of a $^{87}$Rb atom in the case where the quantization axis is $x_3\parallel z$. The fiber radius is $a=180$ nm. The wavelength of the atomic transition is $\lambda_0=516.5$ nm. The refractive indices of the fiber and the vacuum cladding are $n_1=1.4615$ and $n_2=1$, respectively. The field in the fundamental guided mode HE$_{11}$ is quasilinearly polarized along the axis $\xi=x$ (solid lines) or $y$ (dashed lines) and propagates in the direction $f=+1$ or $-1$ of the fiber axis $z$.  The power of the guided light field is $1$ nW. The atom is located on the positive side of the axis $x$ and outside the fiber.}
	\label{fig2}
\end{figure} 

First, we examine the case where the fiber axis $z$ is used as the quantization axis to specify the atomic internal states.
We calculate numerically the absolute value  $|\Omega_q^{(f\xi)}|$ of the Rabi frequency for the quadrupole transition between the sublevel $M=2$ of the hfs level $5S_{1/2}F=2$ and the sublevel $M'=M+q$ of the hfs level $4D_{5/2}F'=4$ of a $^{87}$Rb atom, specified with respect to the quantization axis $x_3\parallel z$. 
We plot in Fig.~\ref{fig2} the radial-distance dependence of $|\Omega_q^{(f\xi)}|$. 
In the calculations for this figure, we have assumed that the field in the fundamental guided mode HE$_{11}$ is quasilinearly polarized along the axis $\xi=x$ (solid lines) or $y$ (dashed lines) and propagates in the positive direction $f=+1$ or the negative direction $f=-1$ of the fiber axis $z$. We observe that $|\Omega_q^{(f\xi)}|$ reduces almost exponentially with increasing $r$. The steep slope in the radial-distance dependence of $|\Omega_q^{(f\xi)}|$ is a signature of the evanescent-wave behavior of the guided field outside the fiber.

Figure \ref{fig2} shows that the absolute value  $|\Omega_q^{(f\xi)}|$ of the Rabi frequency does not depend on the propagation direction $f$ of the field. In other words, the magnitude of the coupling between the field and the atom is symmetric with respect to the opposite propagation directions along the fiber axis. We observe from the dashed curve of Fig.~\ref{fig2}(c) that $\Omega_0^{(fy)}=0$, that is, the atom-field coupling is vanishing  in the case where $q=0$ and $\xi=y$.

\begin{figure}[tbh]
	\begin{center}
		\includegraphics{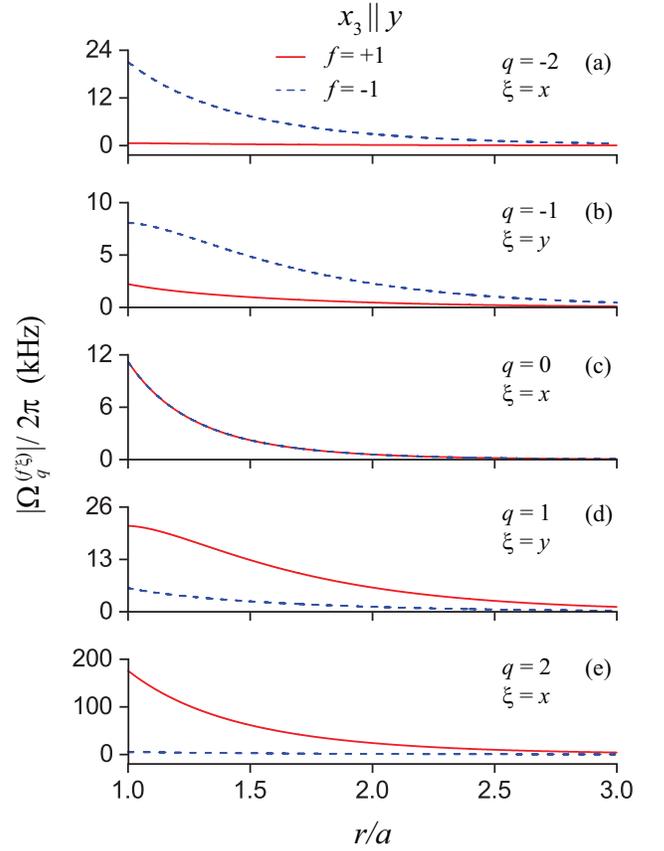}
	\end{center}
	\caption{Radial-distance dependence of the absolute value  $|\Omega_q^{(f\xi)}|$ of the Rabi frequency for the quadrupole transition between the sublevel $M=2$ of the hfs level $5S_{1/2}F=2$ and the sublevel $M'=M+q$ of the hfs level $4D_{5/2}F'=4$ of a $^{87}$Rb atom  in the case where the quantization axis is $x_3\parallel y$. The guided field is quasilinearly polarized along the axis $\xi=x$ in parts (a), (c), and (e) and along the axis $\xi=y$ in parts (b) and (d), and propagates in the positive direction $f=+1$ (solid red lines) or the negative direction $f=-1$ (dashed blue lines) of the fiber axis $z$.  Other parameters are as for Fig.~\ref{fig2}. 
	The Rabi frequencies for  $\xi=y$ and $q=0,\pm2$ and for $\xi=x$ and $q=\pm1$ are vanishing and are therefore not plotted. 
	}
	\label{fig3}
\end{figure} 

Next, we examine the case where the fiber transverse axis $y$ is used as the quantization axis to specify the atomic internal states. We calculate numerically the absolute value  $|\Omega_q^{(f\xi)}|$ of the Rabi frequency for the quadrupole transition between the sublevel $M=2$ of the hfs level $5S_{1/2}F=2$ and the sublevel $M'=M+q$ of the hfs level $4D_{5/2}F'=4$ of a $^{87}$Rb atom, specified with respect to the quantization axis $x_3\parallel y$. 
We plot in Fig.~\ref{fig3} the radial-distance dependence of $|\Omega_q^{(f\xi)}|$. In the calculations for this figure, we have assumed that the guided field is quasilinearly polarized along the axis $\xi=x$ in Figs.~\ref{fig3}(a), \ref{fig3}(c), and \ref{fig3}(e) and along the axis $\xi=y$ in Figs.~\ref{fig3}(b) and \ref{fig3}(d), and propagates in the positive direction $f=+1$ (solid red lines) or the negative direction $f=-1$ (dashed blue lines) of the fiber axis $z$.
The Rabi frequencies for $\xi=y$ and $q=0,\pm2$  and for $\xi=x$ and $q=\pm1$ are  vanishing and are therefore not plotted. The difference between the solid red lines ($f=1$) and the dashed blue lines ($f=-1$) of Fig.~\ref{fig3} shows that the absolute value $|\Omega_q^{(f\xi)}|$ of the Rabi frequency depends on the propagation direction $f$ of the field in the cases of ($\xi=x$, $q=\pm2$) [see Figs.~\ref{fig3}(a) and \ref{fig3}(e)] and  ($\xi=y$, $q=\pm1$) [see Figs.~\ref{fig3}(b) and \ref{fig3}(d)]. We observe from Figs.~\ref{fig3}(a) and \ref{fig3}(b)
that, in the cases of $q=-2$ and $-1$, the value of $|\Omega_q^{(f\xi)}|$ for $f=+1$ (solid red lines) is much smaller than that for $f=-1$ (dashed blue lines). Meanwhile, Figs.~\ref{fig3}(d) and \ref{fig3}(e) show
that,  in the cases of $q=1$ and $2$, the value of $|\Omega_q^{(f\xi)}|$ for $f=+1$ (solid red lines) is much larger than that for $f=-1$ (dashed blue lines). It is clear that the asymmetry of $|\Omega_q^{(f\xi)}|$ with respect to the opposite field propagation directions for $q=\pm2$
is stronger than that for $q=\pm1$. We observe from Fig.~\ref{fig3}(c) that $|\Omega_q^{(f\xi)}|$ does not depend on the field propagation direction $f$ in the case of $q=0$. Comparison between the curves of Fig.~\ref{fig3} shows that the magnitude of $|\Omega_q^{(f\xi)}|$ for the case of ($f=+1$, $q=2$, $\xi=x$) [see the solid red curve of Fig.~\ref{fig3}(e)] is substantially larger than the corresponding values for the other cases of ($f$, $q$, $\xi$).

\begin{figure}[tbh]
	\begin{center}
		\includegraphics{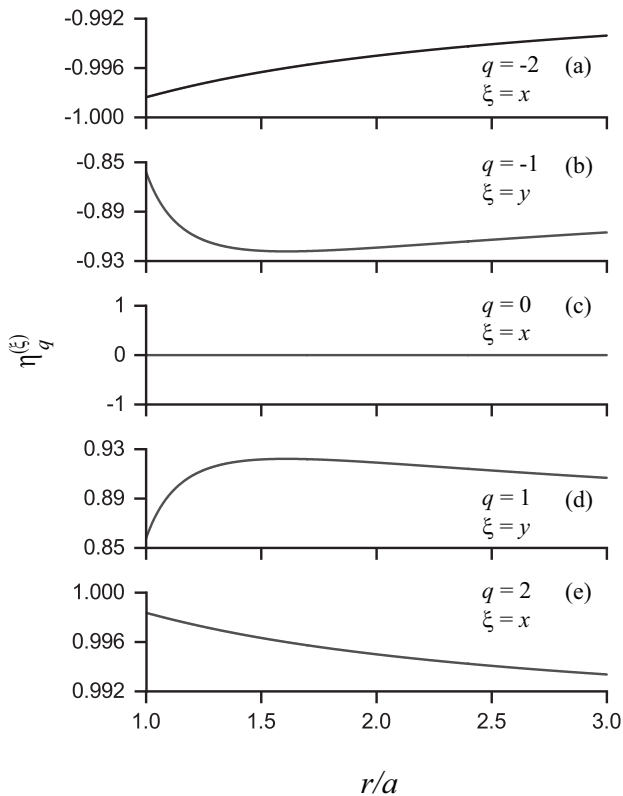}
	\end{center}
	\caption{Asymmetry parameter  $\eta_q^{(\xi)}$ for the directional dependence of the absolute value of the Rabi frequency as a function of the radial distance $r$. The parameters used are as for Fig.~\ref{fig3}. 		
	}
	\label{fig4}
\end{figure} 

The dependence of the absolute value $|\Omega_q^{(f\xi)}|$ of the Rabi frequency on the field propagation direction $f$ is characterized by the asymmetry parameter $\eta_q^{(\xi)}$ [see Eq.~(\ref{q22})]. We plot in Fig.~\ref{fig4} the dependence of $\eta_q^{(\xi)}$ on the radial distance $r$  for the parameters of Fig.~\ref{fig3}. Comparisons between Figs.~\ref{fig4}(a) and \ref{fig4}(e) and between Figs.~\ref{fig4}(b) and \ref{fig4}(d) confirm that $\eta_q^{(\xi)}=-\eta_{-q}^{(\xi)}$. We observe that the asymmetry is vanishing for ($q=0$, $\xi=x$) [see Fig.~\ref{fig4}(c)] but is strong ($|\eta_q^{(\xi)}|>0.85$) for ($q=\pm1$, $\xi=y$) [see Figs.~\ref{fig4}(b) and \ref{fig4}(d)] and very strong ($|\eta_q^{(\xi)}|>0.99$) for ($q=\pm2$, $\xi=x$) [see Figs.~\ref{fig4}(a) and \ref{fig4}(e)]. We also see from Figs.~\ref{fig4}(b) and \ref{fig4}(d) that the absolute value of the factors $\eta_{\pm1}^{(y)}$ has a peak $\max|\eta_{\pm1}^{(y)}|\cong0.92$ at the radial distance $r\cong 1.6a$.
The peak value of the ratio $|\Omega_1^{(+,y)}|/|\Omega_1^{(-,y)}|=|\Omega_{-1}^{(-,y)}|/|\Omega_{-1}^{(+,y)}|$ is about 
$ 4.97$. This value is comparable to the corresponding results for the coupling between an atom with the $\sigma_{\pm}$ dipole transitions and an $x$-polarized guided light field \cite{Mitsch14b,Fam2014}.

\begin{figure}[tbh]
	\begin{center}
		\includegraphics{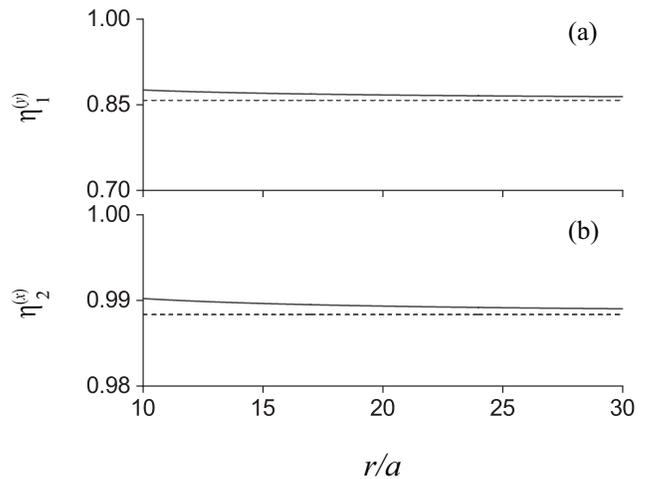}
	\end{center}
	\caption{Asymmetry parameters  $\eta_1^{(y)}$ and $\eta_2^{(x)}$ (solid lines) for the radial distance $r$ in the region $10\leq r/a\leq 30$. The parameters used are as for Fig.~\ref{fig3}. The dotted lines indicate the limiting values $\eta_1^{(y)}(\infty)$ and $\eta_2^{(x)}(\infty)$, calculated from Eqs.~(\ref{q18c}).		
	}
	\label{fig5}
\end{figure} 

According to Eqs.~(\ref{q18c}), the asymmetry parameter $\eta_q^{(\xi)}$ for $q=\pm1$ or $\pm2$ tends to a nonzero value in the limit of large distances $r$. To see this asymptotic behavior, 
we plot in Fig.~\ref{fig5} the radial-distance dependencies of the factors $\eta_1^{(y)}$ and $\eta_2^{(x)}$ (solid lines) in the region $10\leq r/a\leq 30$. The theoretical limiting values $\eta_1^{(y)}(\infty)$ and $\eta_2^{(x)}(\infty)$ [see Eqs.~(\ref{q18c})] are indicated by the horizontal dotted lines. Comparison between the solid and dotted lines shows that the asymptotic behavior of $\eta_1^{(y)}$ and $\eta_2^{(x)}$ agrees very well with the theoretical estimates (\ref{q18c}) for the limiting values.

\begin{figure}[tbh]
	\begin{center}
		\includegraphics{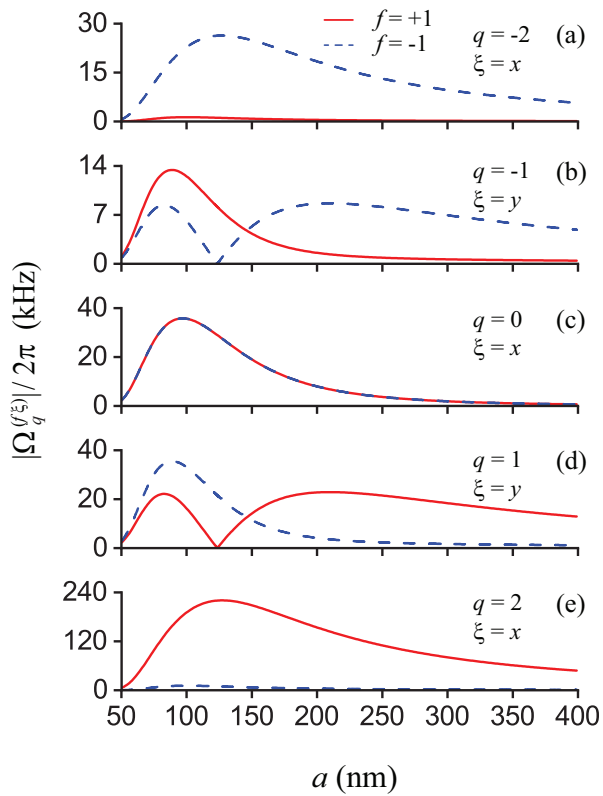}
	\end{center}
	\caption{Absolute value  $|\Omega_q^{(f\xi)}|$ of the Rabi frequency for the quadrupole transition between the sublevel $M=2$ of the hfs level $5S_{1/2}F=2$ and the sublevel $M'=M+q$ of the hfs level $4D_{5/2}F'=4$ of a $^{87}$Rb atom  with the quantization axis $x_3\parallel y$ as a function of the fiber radius $a$. The atom is located on  the fiber surface ($r=a$). Other parameters are as for Fig.~\ref{fig3}. 
	}
	\label{fig6}
\end{figure}

\begin{figure}[tbh]
	\begin{center}
		\includegraphics{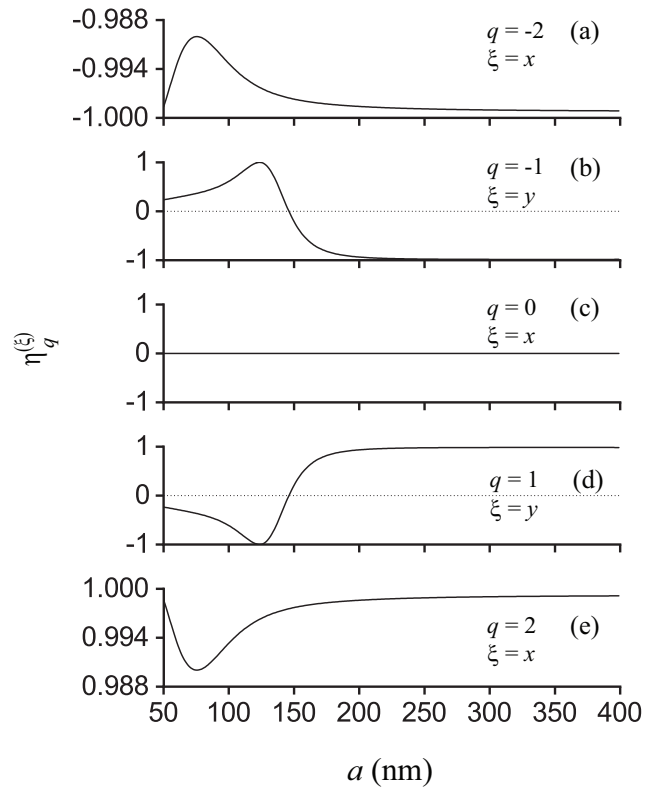}
	\end{center}
	\caption{Asymmetry parameter $\eta_q^{(\xi)}$ for the directional dependence of the absolute value of the Rabi frequency as a function of the fiber radius $a$. The parameters used are as for Fig.~\ref{fig6}. 		
	}
	\label{fig7}
\end{figure}

In order to see the effects of the magnitude of the fiber radius on the direction-dependent coupling between the atom and the guided light field, we plot in Figs.~\ref{fig6} and \ref{fig7} the absolute value  $|\Omega_q^{(f\xi)}|$ of the Rabi frequency and the asymmetry parameter $\eta_q^{(\xi)}$ as functions of the fiber radius $a$. We observe from Fig.~\ref{fig6} that the magnitude of $|\Omega_q^{(f\xi)}|$ for a fixed power has a number of peaks at appropriate values of $a$. Figures \ref{fig6}(a), \ref{fig6}(e), \ref{fig7}(a), and \ref{fig7}(e) show that, for the quadrupole transitions with $q=\pm2$, the asymmetry between the magnitudes $|\Omega_q^{(f\xi)}|$ of the Rabi frequencies for the opposite propagation directions $f=\pm1$  is very strong, namely, $|\eta_q^{(\xi)}|>0.988$.
It is seen from Figs.~\ref{fig6}(b), \ref{fig6}(d), \ref{fig7}(b), and \ref{fig7}(d) that, for the quadrupole transitions with $q=\pm1$, the directional asymmetry between the Rabi frequencies $|\Omega_q^{(f\xi)}|$ varies in a wide range   $1\geq|\eta_q^{(\xi)}|\geq0$. Note that the dashed blue curve in Fig.~\ref{fig6}(b) and the solid red curve in Fig.~\ref{fig6}(d) reach the zero value at  $a\cong 123.5$ nm. This means that the Rabi frequencies $\Omega_{q=-1}^{(f=-1,\xi=y)}$ and $\Omega_{q=1}^{(f=1,\xi=y)}$ become zero when $a\cong 123.5$ nm and $r=a$. For these parameters, the quadrupole transition with $q=-1$ (or $q=+1$) with respect to the quantization axis $y$ is coupled to the forward-propagating (or backward-propagating) $y$-polarized guided light field but not to the corresponding counterpropagating field. Hence, we obtain $|\eta_{\pm1}^{(y)}|=1$ for $a\cong 123.5$ nm [see Figs.~\ref{fig7}(b) and \ref{fig7}(d)]. 
This result indicates that the corresponding quadrupole spontaneous emission into nanofiber-guided modes is unidirectional.
It is worth noting here that, in the region of large fiber radii, the asymmetry parameters $\eta_{\pm1}^{(y)}$ and $\eta_{\pm2}^{(x)}$ approach the limiting values (\ref{q18d}).

\begin{figure}[tbh]
	\begin{center}
		\includegraphics{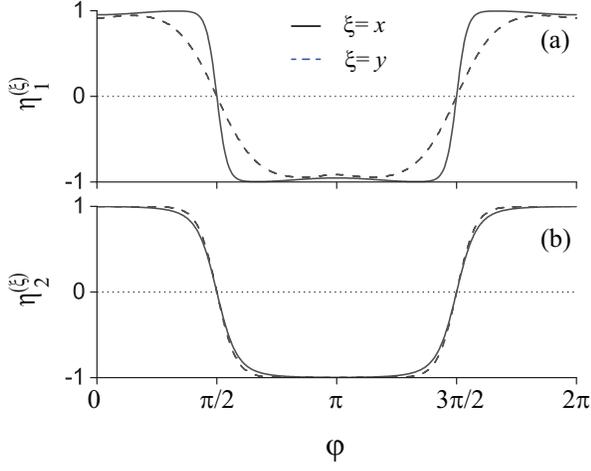}
	\end{center}
	\caption{Asymmetry parameters  $\eta_1^{(\xi)}$ (a) and $\eta_2^{(\xi)}$ (b) as functions of the azimuthal angle $\varphi$ for the position of the atom in the fiber transverse plane $xy$. 
	The quantization axis is $x_3\parallel y$ 
	and the guided field is quasilinearly polarized along the axis $\xi=x$ (solid lines) or $\xi=y$ (dashed lines).
	The radial distance from the atom to the fiber surface is $r-a=50$ nm.
	Other parameters are as for Fig.~\ref{fig3}. 		
	}
	\label{fig8}
\end{figure} 

The asymmetry of the atom-field coupling  depends on the azimuthal position of the atom [see Eqs.~(\ref{q12}) and (\ref{q13})]. To illustrate this fact, we plot in Fig.~\ref{fig8} the asymmetry parameters  $\eta_1^{(\xi)}$ and $\eta_2^{(\xi)}$ as functions of the azimuthal angle  $\varphi$ for the position of the atom in the fiber transverse plane $xy$. We again use the axis $x_3\parallel y$ as the quantization axis. We observe from the figure that the asymmetry is very strong, namely $|\eta_1^{(\xi)}|,|\eta_2^{(\xi)}|\cong 1$, for the positions at $\varphi=0,\pi$. These azimuthal angles correspond to the case of the atom on the $x$ axis, considered in Figs.~\ref{fig3}--\ref{fig7}.  Meanwhile, the asymmetry is vanishing, that is, $|\eta_1^{(\xi)}|,|\eta_2^{(\xi)}|= 0$, for the positions at $\varphi=\pi/2,3\pi/2$, which correspond to the atom on the $y$ axis. Thus, the asymmetry of the atom-field coupling disappears when the radial axis for the atomic position is parallel to the quantization axis, in agreement with Eqs.~(\ref{q24}) and (\ref{q25}).
The solid curve (for $x$-polarized guided light) in Fig.~\ref{fig8}(a) and the dashed curve (for $y$-polarized guided light) in Fig.~\ref{fig8}(b) show that in the limit $\varphi\to0$ or $\pi$, the asymmetry parameters $\eta_1^{(x)}$ and 
$\eta_2^{(y)}$ approach nonzero limiting values although the coupling factors $S_1^{(fx)}$ and $S_2^{(fy)}$ and, hence, 
the Rabi frequencies $\Omega_1^{(fx)}$ and $\Omega_2^{(fy)}$ tend to zero [see the first expression in Eqs.~(\ref{q19}) and the second expression in Eqs.~(\ref{q20})].

\section{Summary}
\label{sec:summary}

In this paper, we have studied the directional dependence of the coupling between a nanofiber-guided light field and a two-level atom with an electric quadrupole transition. We have considered the situation 
where the atom lies on the fiber transverse axis $x$, 
the quantization axis for the atomic internal states is the other orthogonal transverse axis $y$,
the atomic upper and lower levels are the magnetic sublevels $M'$ and $M$ of hfs levels of an alkali-metal atom,
and the field is in a quasilinearly polarized fundamental guided mode HE$_{11}$ with the polarization $\xi=x$ or $y$. 
We have found that the absolute value of the quadrupole Rabi frequency depends on the propagation direction of the light field in the cases of ($M'-M=\pm1$, $\xi=y$) and 
($M'-M=\pm2$, $\xi=x$). This chiral effect occurs as a result of the fact that the strength of the interaction is proportional to a superposition of the gradients of the amplitudes and the spatial phases of the components of the nanofiber-guided field. The directional dependence of the quadrupole Rabi frequency is caused by the contributions originating from either the transverse gradients of the longitudinal field component or the longitudinal gradient of the spatial phase factor. Thus, the directional dependence of the atom-field coupling in the case of quadrupole transitions is not entirely due to spin-orbit coupling of light. We have also found that the directional dependence of the coupling leads to the directional dependence of spontaneous emission into guided modes. 
Our results may open a new way to control and manipulate the interaction between nanofiber-guided light fields and atoms with quadrupole transitions.
 

\begin{acknowledgments}
	This work was supported by the Okinawa Institute of Science and Technology (OIST) Graduate University and by the Japan Society for the Promotion of Science (JSPS) Grant-in-Aid for Scientific Research (C) under Grants No. 19K05316 and No. 20K03795.
\end{acknowledgments}

\appendix

\section{Reduced coupling factors for the atom with the quantization axis $y$}
\label{sec:coupling}

We consider the particular case where the quantization axis is the fiber transverse axis $y$. 
To be concrete, we take $x_1\parallel z$, $x_2\parallel x$, and $x_3\parallel y$. In this case, the reduced coupling factors $S_{M'-M}$ are given by Eqs.~(\ref{q9}). We use the relations
\begin{eqnarray}\label{q10}
	\frac{\partial}{\partial x}&=&\cos\varphi\frac{\partial}{\partial r}
	-\sin\varphi\frac{\partial}{r\partial \varphi},
	\nonumber\\
	\frac{\partial}{\partial y}&=&\sin\varphi\frac{\partial}{\partial r}
	+\cos\varphi\frac{\partial}{r\partial \varphi}.
\end{eqnarray}
Then, for a quasilinearly polarized guided field $\boldsymbol{\mathcal{E}}$ given by Eqs.~(\ref{q7}) and (\ref{q8}), we obtain
\begin{eqnarray}\label{q11}
	&&S_0=
	-\frac{\mathcal{A}}{\sqrt6}\bigg\{ i\beta e_z\cos (\varphi-\varphi_0)
	\nonumber\\&&\mbox{}
	+[e'_r\cos (\varphi-\varphi_0)\cos\varphi
	-ie'_\varphi\sin(\varphi-\varphi_0)\sin\varphi]\cos\varphi
	\nonumber\\&&\mbox{}
	+\frac{1}{r}(e_r+ie_\varphi)\sin(2\varphi-\varphi_0)\sin\varphi
	\nonumber\\&&\mbox{}
	-2[e'_r\cos (\varphi-\varphi_0)\sin\varphi
	+ie'_\varphi\sin (\varphi-\varphi_0)\cos\varphi]\sin\varphi
	\nonumber\\&&\mbox{}
	-\frac{2}{r}(e_r+ie_\varphi)\cos (2\varphi-\varphi_0)\cos\varphi
	\bigg\}e^{if\beta z},
\end{eqnarray}
\begin{eqnarray}\label{q12}
	&&S_{\pm1}=
	\mp\frac{f\mathcal{A}}{2}\bigg\{i\beta[e_r\cos(\varphi-\varphi_0)\sin\varphi
	\nonumber\\&&\mbox{}
	+ie_\varphi\sin(\varphi-\varphi_0)\cos\varphi]
	+e'_z\cos(\varphi-\varphi_0) \sin\varphi
	\nonumber\\&&\mbox{}  
	-\frac{1}{r}e_z\sin(\varphi-\varphi_0)\cos\varphi\bigg\}e^{if\beta z}
	\nonumber\\&&\mbox{}
	+\frac{i\mathcal{A}}{2}\bigg\{ [e'_r\cos(\varphi-\varphi_0)\sin\varphi
	\nonumber\\&&\mbox{}
	+ie'_\varphi\sin (\varphi-\varphi_0)\cos\varphi]\cos\varphi
	\nonumber\\&&\mbox{}
	-\frac{1}{r}(e_r+ie_\varphi)\cos(2\varphi-\varphi_0)\sin\varphi
	\nonumber\\&&\mbox{}
	+[e'_r\cos(\varphi-\varphi_0)\cos\varphi
	-ie'_\varphi\sin(\varphi-\varphi_0)\sin\varphi]\sin\varphi
	\nonumber\\&&\mbox{}	
	-\frac{1}{r}(e_r+ie_\varphi)\sin(2\varphi-\varphi_0)\cos\varphi\bigg\}e^{if\beta z},
\end{eqnarray}
and
\begin{eqnarray}\label{q13}
	&&S_{\pm2}=
	\frac{\mathcal{A}}{2}\bigg\{i\beta e_z\cos (\varphi-\varphi_0)
	\nonumber\\&&\mbox{}	
	-[e'_r\cos (\varphi-\varphi_0)\cos\varphi-ie'_\varphi\sin (\varphi-\varphi_0)\sin\varphi]\cos\varphi
	\nonumber\\&&\mbox{}
	-\frac{1}{r}(e_r+ie_\varphi)\sin(2\varphi-\varphi_0)\sin\varphi\bigg\}e^{if\beta z}
	\nonumber\\&&\mbox{}
	\mp\frac{if\mathcal{A}}{2}\bigg\{i\beta[e_r\cos (\varphi-\varphi_0)\cos\varphi-ie_\varphi\sin (\varphi-\varphi_0)\sin\varphi]
	\nonumber\\&&\mbox{}
	+e'_z\cos (\varphi-\varphi_0)\cos\varphi+\frac{1}{r}e_z\sin(\varphi-\varphi_0) \sin\varphi\bigg\}e^{if\beta z}.
	\nonumber\\
\end{eqnarray}

\subsection{Atom on the $x$ axis}

We assume that the atom is located on the positive side of the axis $x$, that is, $\varphi=0$. In this case, we have
\begin{eqnarray}\label{q14}
	S_0&=&
	-\frac{\mathcal{A}}{\sqrt6}\bigg[i\beta e_z+e'_r
	-\frac{2}{r}(e_r+ie_\varphi)\bigg]\cos\varphi_0 \; e^{if\beta z},
	\nonumber\\
	S_{\pm1}&=&
	\frac{\mathcal{A}}{2}\bigg[\mp f\bigg(\beta e_\varphi  
	+\frac{1}{r}e_z\bigg)+e'_\varphi
	+\frac{i}{r}(e_r+ie_\varphi)\bigg]
	\nonumber\\&&\mbox{}
	\times\sin\varphi_0 \; e^{if\beta z},
	\nonumber\\
	S_{\pm2}&=&
	\frac{\mathcal{A}}{2}[i\beta e_z-e'_r
	\mp f(ie'_z-\beta e_r)]\cos\varphi_0 \;	e^{if\beta z}. \qquad
\end{eqnarray}

When the guided field is polarized along the $x$ direction, we have $\varphi_0=0$. In this case, Eqs.~(\ref{q14}) yield
\begin{eqnarray}\label{q17}
	S_0&=&
	-\frac{\mathcal{A}}{\sqrt6}\bigg[i\beta e_z+e'_r
	-\frac{2}{r}(e_r+ie_\varphi)\bigg]e^{if\beta z},
	\nonumber\\
	S_{\pm1}&=&0,
	\nonumber\\
	S_{\pm2}&=&
	\frac{\mathcal{A}}{2}[i\beta e_z-e'_r
	\mp f(ie'_z-\beta e_r)]e^{if\beta z}.\qquad
\end{eqnarray}
With the help of Eqs. (\ref{7a}), we can show that $|S_{\pm2}|$ has different values for different $f$.

When the guided field is polarized along the $y$ direction, we have $\varphi_0=\pi/2$. In this case, Eqs.~(\ref{q14}) yield
\begin{eqnarray}\label{q18}
	S_0&=&0,
	\nonumber\\
	S_{\pm1}&=&
	\frac{\mathcal{A}}{2}\bigg[\mp f\bigg(\beta e_\varphi  
	+\frac{1}{r}e_z\bigg)+e'_\varphi
	+\frac{i}{r}(e_r+ie_\varphi)\bigg]e^{if\beta z},
	\nonumber\\
	S_{\pm2}&=&0.
\end{eqnarray}
With the help of Eqs. (\ref{7a}), we can show that $|S_{\pm1}|$ depends on $f$.

\subsection{Atom on the $y$ axis}

We now assume that the atom is located on the positive side of the axis $y$, that is, $\varphi=\pi/2$. In this case, we have
\begin{eqnarray}\label{q21}
	S_0&=&
	-\frac{\mathcal{A}}{\sqrt6}\bigg[i\beta e_z+\frac{1}{r}(e_r+ie_\varphi)-2e'_r
	\bigg]\sin\varphi_0 \; e^{if\beta z},
	\nonumber\\
	S_{\pm1}&=&
	\frac{\mathcal{A}}{2}\bigg\{\mp f(i\beta e_r+e'_z)\sin\varphi_0
	\nonumber\\&&\mbox{}
	+\bigg[\frac{i}{r}(e_r+ie_\varphi)
	+e'_\varphi\bigg]\cos\varphi_0\bigg\}e^{if\beta z},
	\nonumber\\
	S_{\pm2}&=&
	\frac{\mathcal{A}}{2}\bigg\{\bigg[i\beta e_z	
	-\frac{1}{r}(e_r+ie_\varphi)\bigg]\sin\varphi_0
	\nonumber\\&&\mbox{}
	\mp if\bigg(\beta e_\varphi
	+\frac{1}{r}e_z\bigg)\cos\varphi_0 \bigg\}e^{if\beta z}.\qquad
\end{eqnarray}
For the $x$-polarized guided field (with $\varphi_0=0$), we obtain
\begin{eqnarray}\label{q24}
	S_0&=&0,\nonumber\\
	S_{\pm1}&=&
	\frac{\mathcal{A}}{2} \bigg[\frac{i}{r}(e_r+ie_\varphi)+e'_\varphi\bigg] e^{if\beta z},\nonumber\\
	S_{\pm2}&=&
	\mp \frac{if\mathcal{A}}{2}
	\bigg(\beta e_\varphi+\frac{1}{r}e_z\bigg)e^{if\beta z}.
\end{eqnarray}
For the $y$-polarized guided field (with $\varphi_0=\pi/2$), we get
\begin{eqnarray}\label{q25}
	S_0&=&
	-\frac{\mathcal{A}}{\sqrt6}\bigg[i\beta e_z+\frac{1}{r}(e_r+ie_\varphi)-2e'_r
	\bigg] e^{if\beta z},\nonumber\\
	S_{\pm1}&=&
	\mp\frac{f\mathcal{A}}{2} (i\beta e_r+e'_z) e^{if\beta z},\nonumber\\
	S_{\pm2}&=&
	\frac{\mathcal{A}}{2}\bigg[i\beta e_z	
	-\frac{1}{r}(e_r+ie_\varphi)\bigg]e^{if\beta z}.
\end{eqnarray}
It is clear that, for both $x$- and $y$-polarized guided light fields, the absolute values of the coupling factors $S_q$ do not depend on the propagation direction $f$.

\section{Quadrupole spontaneous emission of the atom into nanofiber-guided modes}
\label{sec:spont}

We consider the electric quadrupole interaction between the atom and the quantum nanofiber-guided field.
We assume that the fiber supports only the fundamental guided mode HE$_{11}$  \cite{fiber books} 
in a finite bandwidth around the central frequency $\omega_0$ of the atom. 
We label each guided mode in this bandwidth by an index $\mu=(\omega,f,\xi)$. 
Here, $\omega$ is the mode frequency, $f=+1$ or $-1$ denotes the forward or backward propagation direction along the fiber axis $z$, and $\xi=x$ or $y$ is the $x$ or $y$ quasilinear polarization. We neglect the effects of the radiation modes \cite{fiber books}.

In the interaction picture, the quantum expression for the positive-frequency part $\mathbf{E}^{(+)}_{\mathrm{g}}$ of the electric component of the field in the guided modes is \cite{cesium decay}
\begin{equation}
	\mathbf{E}^{(+)}_{\mathrm{g}}=i\sum_{\mu}\sqrt{\frac{\hbar\omega\beta'}{4\pi\epsilon_0}}
	\;a_{\mu}\mathbf{e}^{(\mu)}e^{-i\omega t}.
	\label{3}
\end{equation}
Here, $\mathbf{e}^{(\mu)}=\mathbf{e}^{(\mu)}(r,\varphi,z)$ is the normalized profile function of the guided mode $\mu$ in the classical problem, $a_{\mu}$ is the corresponding photon annihilation  operator, 
$\sum_{\mu}=\sum_{f\xi}\int_0^{\infty}d\omega$ is the generalized summation over the guided modes,
$\beta$ is the longitudinal propagation constant, and $\beta'$ is the derivative of $\beta$
with respect to $\omega$. 
The mode profile functions $\mathbf{e}^{(\omega f\xi)}$ for the quasilinear polarizations $\xi=x$ and $y$ are given as
\begin{eqnarray}\label{4}
	\mathbf{e}^{(\omega fx)}&=&\mathcal{A} (\hat{\mathbf{r}}e_r\cos\varphi
	+i\hat{\boldsymbol{\varphi}}e_\varphi\sin\varphi 	
	+f\hat{\mathbf{z}}e_z\cos\varphi) e^{if\beta z},\nonumber\\
	\mathbf{e}^{(\omega fy)}&=&\mathcal{A} (\hat{\mathbf{r}}e_r\sin\varphi
	-i\hat{\boldsymbol{\varphi}}e_\varphi\cos\varphi 	
	+f\hat{\mathbf{z}}e_z\sin\varphi) e^{if\beta z}.\nonumber\\
\end{eqnarray}
The normalization condition 
\begin{equation}\label{5}
	\int _{0}^{2\pi}d\varphi\int _{0}^{\infty}n_{\mathrm{ref}}^2\,|\mathbf{e}^{(\mu)}|^2r\,dr=1
\end{equation}
is required, where $n_{\mathrm{ref}}(r)=n_1$ for $r<a$ and $n_2$ for $r>a$.
The operators $a_{\mu}$ and $a_{\mu}^\dagger$ satisfy the continuous-mode bosonic commutation rules $[a_{\mu},a_{\mu'}^\dagger]=\delta(\omega-\omega')\delta_{ff'}\delta_{\xi\xi'}$. 

Assume that the atom is positioned at a point $(r,\varphi,z)$ outside the fiber. In the interaction picture, the Hamiltonian for the electric quadrupole interaction between the atom and the quantum guided field in the rotating-wave approximation is given by 
\begin{equation}\label{6}
	H_{\mathrm{int}}=-i\hbar\sum_{\mu}G_{\mu}
	\sigma_{eg}a_{\mu}e^{-i(\omega-\omega_0)t}+\mbox{H.c.},
\end{equation}
where the coefficients 
\begin{eqnarray}\label{7}
	G_{\mu}=\frac{1}{12}\sqrt{\frac{\omega\beta'}{\pi\epsilon_0\hbar}}
	\sum_{ij}\langle e|Q_{ij}|g\rangle\frac{\partial e^{(\mu)}_j}{\partial x_i}(0)
\end{eqnarray}
characterize the coupling between the atom  and the guided mode $\mu$.
Inserting Eq.~(\ref{q2}) into Eq.~(\ref{7}) yields 
\begin{eqnarray}\label{8}
	G_{\mu}&=&\sqrt{\frac{\hbar\omega\beta'}{4\pi\epsilon_0}}\;C_{F'M'FM}S^{(\mu)}_{M'-M},
\end{eqnarray}
where
\begin{equation}\label{10}
	S^{(\mu)}_{M'-M}=S_{M'-M}|_{\boldsymbol{\mathcal{E}}=\mathbf{e}^{(\mu)}}=\sum_{ij}u_{ij}^{(M'-M)}\frac{\partial e^{(\mu)}_j}{\partial x_i}(0)
\end{equation}
is the reduced coupling factor for the normalized field $\mathbf{e}^{(\mu)}$. 
The coefficient $C_{F'M'FM}$ in Eq.~(\ref{8}) is given by Eq.~(\ref{q6a}).

We use the Fermi golden rule \cite{Loudon} to calculate
the rate $\gamma_{\mathrm{g}}^{(f\xi)}$ of spontaneous emission from the atom into the nanofiber-guided modes
with the propagation direction $f$ and the polarization $\xi$. 
We find 
\begin{equation}\label{11}
	\gamma_{\mathrm{g}}^{(f\xi)}=2\pi|G_{\omega_0f\xi}|^2
	=\frac{\hbar\omega_0\beta_0'}{2\epsilon_0}\;|C_{F'M'FM}|^2|S^{(\mu_0)}_{M'-M}|^2,
\end{equation}
where $\mu_0=(\omega_0f\xi)$ and $\beta_0'=\beta'(\omega_0)$.
It is clear that $\gamma_{\mathrm{g}}^{(f\xi)}$ is proportional to  $|S_{M'-M}^{(\mu_0)}|^2$.


\end{document}